\documentclass[10pt,letterpaper]{article}
\usepackage[top=0.85in,left=2.75in,footskip=0.75in,marginparwidth=2in]{geometry}

\usepackage[utf8]{inputenc}

\usepackage[style=nature]{biblatex}
\usepackage{nameref,hyperref}

\usepackage[right]{lineno}

\usepackage{microtype}
\DisableLigatures[f]{encoding = *, family = * }

\usepackage{changepage}

\usepackage[aboveskip=1pt,labelfont=bf,labelsep=period,singlelinecheck=off]{caption}

\usepackage{todonotes}

\makeatletter
\renewcommand{\@biblabel}[1]{\quad#1.}
\makeatother

\usepackage{lastpage,fancyhdr,graphicx}
\usepackage{epstopdf}
\fancyheadoffset[L]{2.25in}
\fancyfootoffset[L]{2.25in}

\usepackage{color}

\definecolor{Gray}{gray}{.25}

\usepackage{graphicx}

\usepackage{wrapfig}
\usepackage[pscoord]{eso-pic}
\usepackage[fulladjust]{marginnote}
\reversemarginpar

\begin{document}

\begin{flushleft}
{\Large
\textbf\newline{cellSTORM – Cost-effective Super-Resolution on a Cellphone using dSTORM}
}
\newline
Benedict Diederich\textsuperscript{1,2,*},
Patrick Then\textsuperscript{1,2}
Alexander Jügler\textsuperscript{1,2},
Ronny Förster\textsuperscript{1,2},
Rainer Heintzmann\textsuperscript{1,2}

\bigskip
\textbf{1} Leibniz Institute of Photonic Technology, Albert-Einstein Str. 9, 07745 Jena, Germany
\\
\textbf{2} Institute of Physical Chemistry and Abbe Center of Photonics, Friedrich-Schiller-University Jena, Helmholtzweg 4, 07743 Jena, Germany
\\
\bigskip
* benedict.diederich@ipht-jena.de

\end{flushleft}

\section*{Abstract}

Expensive scientific camera hardware is amongst the main cost factors in modern, high-performance microscopes. On the other hand, cheap, consumer-grade camera devices can provide surprisingly good performance. Widely available smartphones include cameras, providing a good opportunity for "imaging on a budget". Yet, Single-Molecule-Localization-Microscopy (SMLM) techniques like Photoactivated Localization Microscopy (PALM) or (direct) Stochastic Optical Reconstruction Microscopy \textit{d}STORM, are demanding in terms of photon sensitivity and readout noise, seemingly requiring a scientific-grade camera. Here we show that super-resolution imaging by \textit{d}STORM is possible using a consumer grade cellphone camera. Trained image-to-image generative adversarial network (GAN), successfully improves the signal-to-noise ratio (SNR) by compensating noise and compression artifacts in the acquired video-stream at poor imaging conditions. 

We believe that "cellSTORM" paves the way for affordable super-resolution microscopy
suitable for research and education. Our low-cost setup achieves optical resolution below 80\,nm yielding wide access to cutting edge research to a big community.


\section*{Introduction}

Super-resolution by Single-Molecule-Localization-Microscopy (SMLM) techniques like PALM (Photoactivated-Localization Microscopy) \cite{Betzig2006, Hess2006} or \textit{d}STORM \cite{Rust2006} is well established in biology and medical research. Together with other modalities like STED \cite{Hell1994, Klar1999} and SIM \cite{Heintzmann2017, Tolstik2015} SMLM revolutionized optical far-field microscopy beyond the Abbe limit \cite{Abbe1873}. 
\newline 
Typically these methods are connected with costly hardware for excitation and detection \cite{Saurabh2012, Diekmann}. Scientific grade sCMOS or emCCD cameras are a major cost factor, as high photon efficiency and low noise are paramount.

However, the development of mobile phones created surprisingly powerful cameras and sensors, worth considering. \newline

Almost 2.3 Billion smartphones are actively in use \cite{Statista2017}. The effect is the availability of high quality cameras with big computational power. All that at a very low price, where each development iteration aims to further optimize not only the image quality, but also the general performance of the handheld devices \cite{Dendere2015}.
\newline
This gave rise to the development of mobile microscopy, which so far resulted e.g. in hand-held devices capable of quantitative phase-imaging of biological material by combining the cellphone with customized hardware adapters\cite{Dong2014, Breslauer2008, Phillips2015a} or act as portable diagnosis devices to detect e.g. waterborne parasites \cite{Phillips2015a, Rasooly2016, Diederich2017c, Phillips2015b, Orth2018}. 

Successful attempts to image single molecules using low-cost CMOS cameras to build cost-effective SMLM setups e.g. for educational environments \cite{Ma2017, Holm2014, Diekmann} demonstrate that the imaging quality of such sensors suffices for SMLM. 
The small footprint and availability makes the technology suitable for educational or high bio-safety-level lab environments, as well as rapid-prototyping applications, where the processing can be done on the cellphone itself to keep size and costs small. \newline

Commonly smartphone camera sensors, are equipped with bayer patterned filters, significantly lowering the detetion efficiency compared to monochromatic imaging. Recently, some new camera modules, as in the HUAWEI P9, possess sensitive monochromatic CMOS camera chips. However, acquiring high quality RAW data using a cellphone is still very challenging. Simplified hardware abstraction layers embedded into the firmware of the camera module prevent accessing the raw pixel values. Compression and noise artifacts are therefore a potential problem of imaging with smartphone cameras.
 
Nevertheless, a simple adaption of the mobile phone device a common brightfield microscope equipped with an excitation laser combination with tailored image processing algorithms should allow to gain super-resolution information well below the diffraction limit. 
\newline

First, we investigate the image formation process of the smartphone camera and calibrate its noise performance. We then investigate the recovery of lost information based on machine learning approaches before we compare superresolved localization images acquired with a professional emCCD camera and a smartphone.

\section{Materials and Methods}

\subsection{Optical setup}

The basic \textit{d}STORM-system is realized with a standard inverted microscope stand (AxioVert 135 TV, Zeiss, Germany) equipped with a nosepiece-stage (IX2-NPS, Olympus, Japan) to keep drift low. A  637\,nm diode laser (P=150 mW, OBIS, Coherent, USA) is focused to the back-focal plane of the microscope objective lens (ZEISS $100\,\times$, NA=1.46) to realize a homogenous illumination in the sample plane. Using an adjustable mirror, it is also possible to change the laser position in the back-focal plane. This enables background-free total internal reflection (TIRF) illumination. 
\newline

An emCCD camera (iXon3 DU-897, Andor, UK, Table \ref{tab1}) can be used to image the sample in widefield and STORM-mode during normal operation. For imaging via the cellphone, the beam-path is switched from the camera port to the eyepiece, where a common $10\,\times$ monocular eyepiece is equipped with a custom-made 3D-printed cellphone adapter \cite{OpenOcular2017}. The cellphone (P9 EVA-L09, HUAWEI, China, Table \ref{tab1}) is placed with its camera lens in the Ramsden disk of the eyepiece (see Fig \ref{fig:fig1} in Supplement \ref{sec:hardset}), since an eyepiece images the intermediate image produced by the tube lens of the Axiovert body to infinity. Correct imaging is assured, when the exit-pupil of the microscope is sharply imaged on the cellphone’s sensor. To keep distortions like field curvature or vignetting as low as possible, we cropped the video-frames to an active area of $512\times512$ pixels around the center of the frame. Additionally the focus of the photo-lens was fixed to infinity to minimize any extra aberrations. 
\newline

\begin{table}[!ht]
	\caption[Tab 1: ]{Comparison of a scientific-grade with a low-cost cell phone camera}
	\begin{tabular}{l|c|c}
		\hline 
		& \textbf{Andor iXonEM+ 897 }& \textbf{HUAWEI P9 (EVA-L09,} \\ 
		&  &  \textbf{Sony IMX286, Grayscale } \\ 
		\hline 
		Pixel\#: & $512\times 512$ &  $3980\times 2460$ \\ 
		\hline 
		Sensortype &(back-illuminated) emCCD & (back-illuminated) CMOS\\ 
		\hline 
		Pixelsize ($\mu$m): & 16  & 1,25 \\ 
		\hline 
		Bitdepth: & 14 Bit&  12 Bit  \\ 
		\hline 
		Read Noise ($e^-$): & 0,2 RMS  &  1,23 RMS (see Supplement \ref{sec:camchar}) \\ 
		\hline 
		Quantum-Efficiency & $\geq90\%$ & $\approx70-80\% $ \\ 
		\hline 
	\end{tabular} 
	\label{tab1}
\end{table}

We use the 12 Bit monochromatic sensor chip (Sony IMX 286, Japan, Table \ref{tab1}, \cite{Inc.b} of the P9’s dual-camera module. 

\subsection{Image acquisition and reconstruction}
The aim of the camera manufacturers is to ensure optimal image quality in everyday environments. Tailored algorithms help to hide problems introduced e.g. by the small pixels and lens dimensions \cite{Galstian, Nakamura}. 

In contrast to industry-standard CMOS cameras, which mostly rely on the same sensor types, the entire post-processing of the acquired images is done by proprietary firmware on a dedicated circuit, called image signal processor (ISP) \cite{QualcommTechnologies2014, SonyCoperation2014, Ova}. This allows real-time optimization of the image quality. It is mainly responsible for demosaicing the Bayer-pattern to generate RGB images, but also reduces the effect of lens aberrations and removes hot-pixels or thermal noise \cite{BrianKlug2013, Galstian}. The ISP takes also care of the hardware control (e.g. autofocus, optical image stabilization) and encoding of the video-stream into less memory consuming formats \cite{GoogleInc.2015}. The smartphone camera is optimized for beautiful rather than accurate images, thus the scientific relevant information for SMLM of the position dependent quantity of the photons is partially lost by the on-chip processing.

\subsection{Cellphone data acquisition}
\label{sec:celldataacq}

Cellphones of the newer generations allow to access the raw camera sensor pixel values, i.e. the sensor data before further processing or compression by e.g. JPEG/MPEG algorithms. Recording the unprocessed raw-data ("snap-mode") makes the use of cellphones even more attractive for science \cite{Diederich2017c, Smith2011b, Orth2018, Skandarajah2014a, Sung2017}. 

However, \textit{d}STORM requires continuous (to keep the molecules in the dark state) and fast acquisition over several minutes which is incompatible with snap-mode acquisition. The computational effort to save a time-series of raw-frames makes it impracticable for these measurements.

Hence, we were forced to use the standard time-series acquisition mode ("video-mode"), where the compression of the raw data was unavoidable.

Acquiring monochromatic video-sequences is not part of the cellphone’s software by default. To circumvent this issue, we wrote a customized APP (application) based on the Google Android Camera2 API \cite{GoogleInc.2015} and an open-sourced camera library "FreeDcam" \cite{Fuchs2018, Diederich2018}, which enables the full control over the camera parameter like sensitivity (ISO), focus position, exposure time and frame-rate, as well as the access to the monochromatic chip. 

Following the Camera2 API, the camera acts as a server which receives capture requests. These encompass hardware settings of an individual frame or sequences of images. To capture monochromatic images, the vendor-specific settings have to be accessed, hidden in the JAVA reflection classes. Once a camera request is processed, the camera yields a capture result, which has to be drawn on a surface object and can later be processed by a media recorder. In case of an Android device, the latter can be either a JPEG or DNG writer, both of which rely on hardware accelerated processing using java native interface (JNI).
\newline
Down-converting the video-stream, e.g. using the H264 video-codec, is also implemented on the ISP. To reduce the amount of memory, it relies on the exact-match integer transform \cite{ITU-T2013, Wiegand2003, Nakamura} which uses reference images and calculates residual/difference images to reduce the amount of redundant information. The lossy compression incorporates accurate information of the actual pixel-values, necessary for precise localization of the fluorophores according to our experiments.

\subsubsection{Neuronal networks for image enhancement }
\label{sec:neuralnetwork}
In order to reduce imaging artifacts resulting from high pixel-noise or strong compression (see Supplement \ref{sec:hardset}), we made an implicit black-box model (an artificial neural network) of the camera, describing the image degradation process with the aim to recover the image information prior to compression from the measured data.
\newline 

Machine learning can learn an implicit model which maps a set of input variables onto a set of outputs \cite{Cicek2016, Ronneberger2015, Nehme2018, Creswell}. A large variety of different network architectures have shown, that, resulting from prior knowledge, single-image super-resolution (interpolation) \cite{Glasner2009} or recovering the optical phase from an intensity image \cite{Sinha2017} is possible. The popular image-to-image GAN (pix2pix) \cite{Isola2016b} generates images by learning features from a set of inputs and applies it to the outputs (e.g. colorizing b/w image). Compared to former approaches, based on encoder-decoder networks \cite{Johnson2016}, the cost-function for the training process is not directly defined but trained alongside the generator. This enables a customized cost-function which, fits the model better and avoids issues of blurring the generated results for example in case of the L2-norm \cite{Isola2016b, Wang}. 
\newline
Previous approaches trained a noise model using adversarial network architectures (GAN) \cite{Goodfellow2014, Creswell} or directly localized STORM events using an auto-encoder \cite{Nehme2018} or Baysian-statistics \cite{Boyd2018}.   
\newline

Here we aimed to enhance the image quality, after the raw frames of point-emitters were altered by the unknown image preprocessing and compression of the HUAWEI P9 cellphone. Following a modified version of the pix2pix network architecture (see Supplement \ref{sec:nnarchitecture}). The goal of the network is to recover high quality (e.g. high SNR) images (reffered to "\textit{B}") from degraded video-frames of real STORM measurements (referred to "\textit{A}"). To produce training datasets, we developed three different methods to generate “A to B” data-pairs. 

The first approach is based on purely simulated data, where blinking events are degraded based on our camera model described in in Supplement \ref{sec:camchar}.
A second approach holds simulated \textit{d}STORM events with known positions displayed on a secondary screen and their recorded frames from the cellphone camera. 
The third method takes a localized result from a real \textit{d}STORM measurement from a cellphone and produces perfect PSFs for each frame and combines it with the noisy video frames. 
\newline
It turned out, that mixing data based on the first and third method gives the best results. Further details on the generation of the dataset can be found in Supplement \ref{sec:nndataset}.

\subsection{dSTORM imaging samples}
Cell samples have been prepared from HeLa cells using the PFA-fixation protocol outlined in \cite{Whelan2015}. Microtubuli have been stained using Monoclonal mouse anti-$\beta$ -tubulin (Sigma Aldrich) and Goat anti-Mouse IgG secondary antibody (ThermoFisher Scientific), labeled with Alexa Fluor 647 at 1:150 and 1:300 dilution, respectively. All imaging experiments have been conducted in imaging buffer prepared freshly from 150-200mM MEA ($\beta$-Mercapto-ethylamine hydrochloride) in PBS and pH adjusted to 7.4 using NaOH. The oxygen scavenging effect from MEA has been proven efficient enough to refrain from additional enzyme-based oxygen scavenger systems. 

\section{Results}

We divide the results into two parts. First, we show our localized SMLM data using a smartphone under optimal imaging conditions which are directly suitable for standard \textit{d}STORM image processing and then we present a second less optimal dataset, where preprocessing by a NN was required to yield results of good quality. \newline
The less optimal data set should simulate the use of cameras with less quality of not optimal chosen parameters for the acquisition process. 

\subsection{Localization with Compressed Smartphone Data}

The reduced SNR (signal-to-noise ratio) and additional artifacts from video compression are directly apparent in the image sequences acquired by a cellphone camera. Nevertheless, the robust nature of the reconstruction algorithms enables successful localization of blinking events even under non-ideal conditions. Here, we used the freely available rapidSTORM \cite{Wolter2012} and ThunderSTORM \cite{Ovesny2014} software, which yielded comparable results when reconstructing the final image from recorded data. 
\newline

\begin{figure}[h!]
	\centering
	\includegraphics[width=1\linewidth]{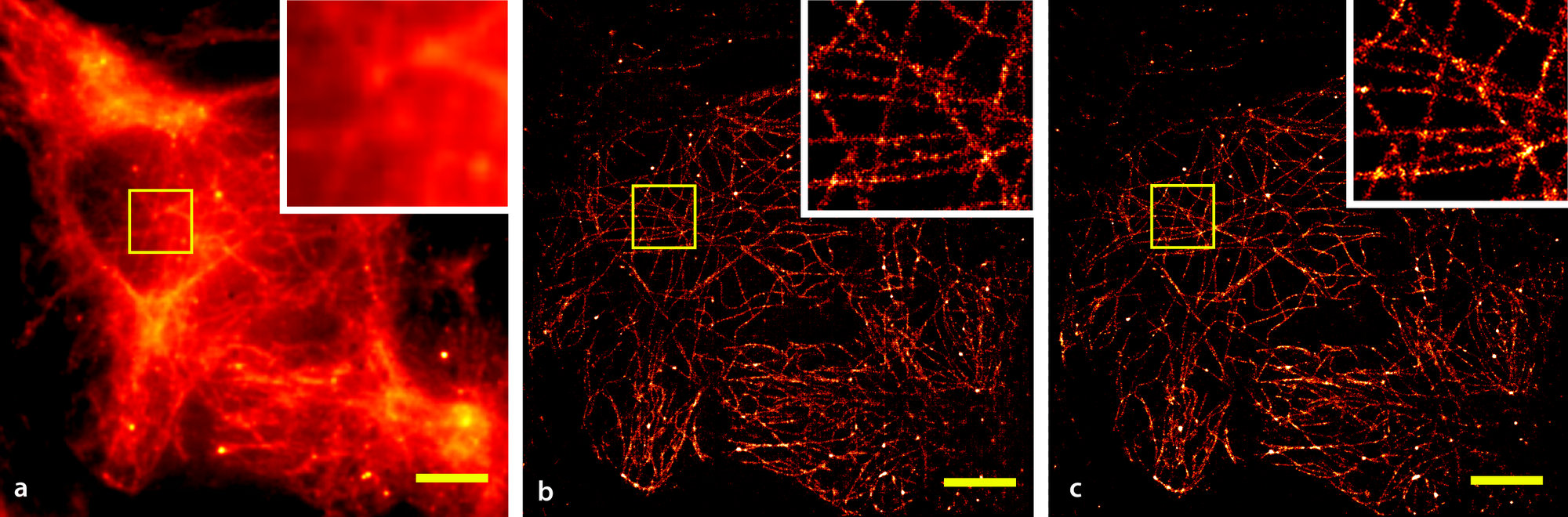}
	\caption[Fig 3:]{Microtubuli in HeLa cells stained using AlexaFluor 647 labeled antibodies and recorded with the cellphone camera after reconstruction with rapidSTORM (a) and a picture obtained through summing all recorded frames (b), equivalent to a widefield image. No drift correction was applied to the time-series. (Scalebar = $1\,\mu m$)}
	\label{fig:fig3}
\end{figure}

Figure \ref{fig:fig3} shows the results obtained with cellSTORM from HeLa cells stained for tubulin using AlexaFluor 647-labeled primary/secondary antibodies. After applying thunderSTORM directly to the $\approx$18 thousand acquired video images (at 20\,fps), the structure of microtubuli is clearly resolved at a resolution of 75\,nm measured using the Fourier ring correlation \cite{Banterle2013}. To compare to conventional \textit{d}STORM data, we recorded another series of a similar cell using the emCCD camera of our setup. Due to the already low photon yield at the cellphone camera, we opted against using an beamsplitter to simultaneously record the same area with cellphone and emCCD camera and instead imaged separate cells of the same sample.   

\begin{figure}[h!]
	\centering
	\includegraphics[width=1\linewidth]{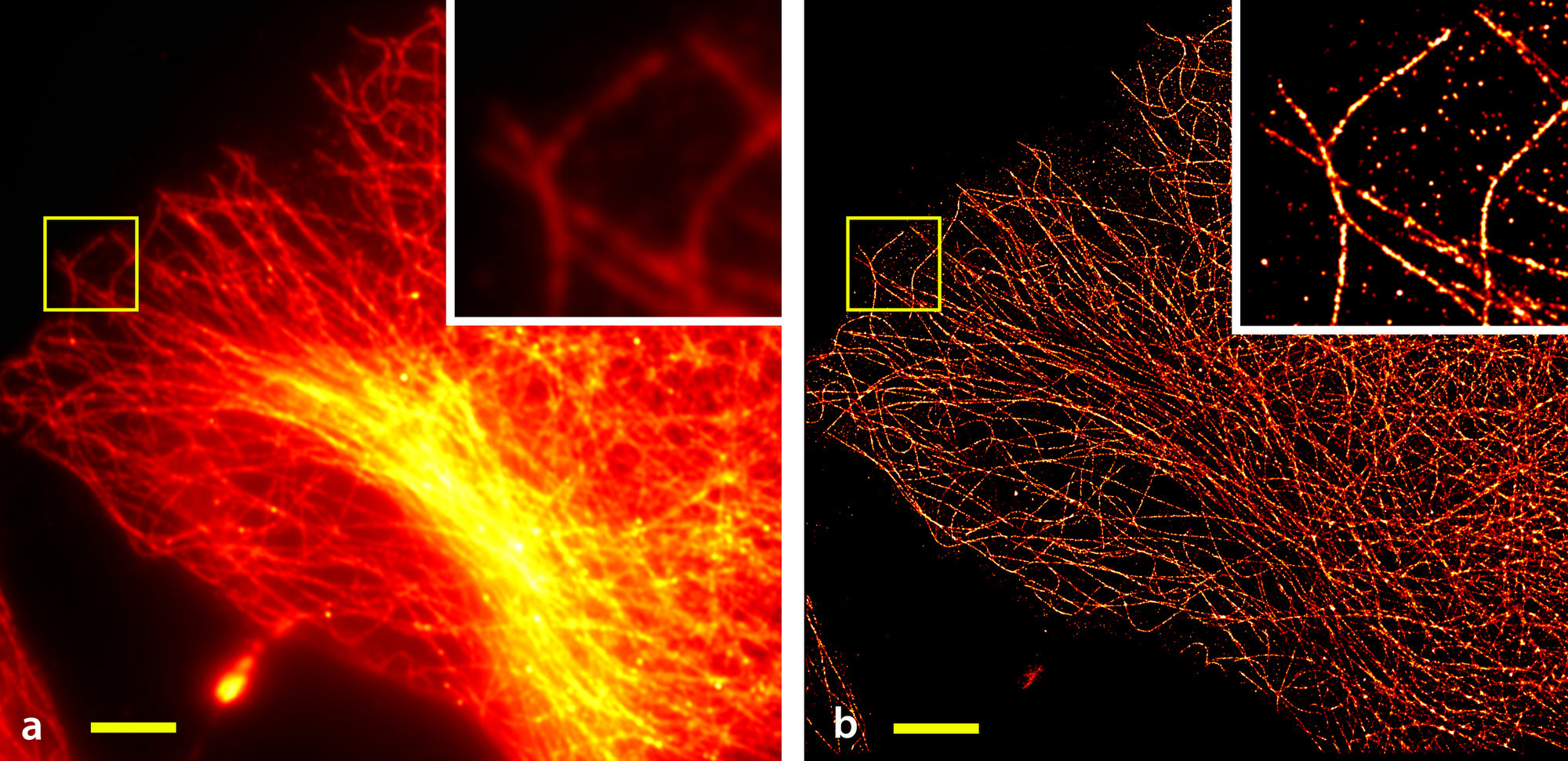}
	\caption[Fig 3:]{Microtubuli in HeLa cells stained using AlexaFluor 647 labeled antibodies and recorded with the Andor emCCD camera after reconstruction with rapidSTORM (a) and a picture obtained through summing all recorded frames (b), equivalent to a widefield image. No drift correction was applied to the time-series. (Scalebar = $10\,\mu m$)}
	\label{fig:fig2a}
\end{figure}

The images acquired using the professional emCCD camera \ref{fig:fig2a} under identical buffer and illumination conditions yielded a final resolution of 45\,nm. While this number is smaller than for cellSTORM the relative difference is nonetheless surprisingly small. 

\subsection{Localization with post-processing}
Especially in poor imaging condition, applying our learned black-box model described in section \ref{sec:neuralnetwork} turned out to be beneficial. Each frame using the trained network, before the processed image-stack is fed into the localization software. 
\newline

\begin{figure}[h!]
	\centering
	\includegraphics[width=1\linewidth]{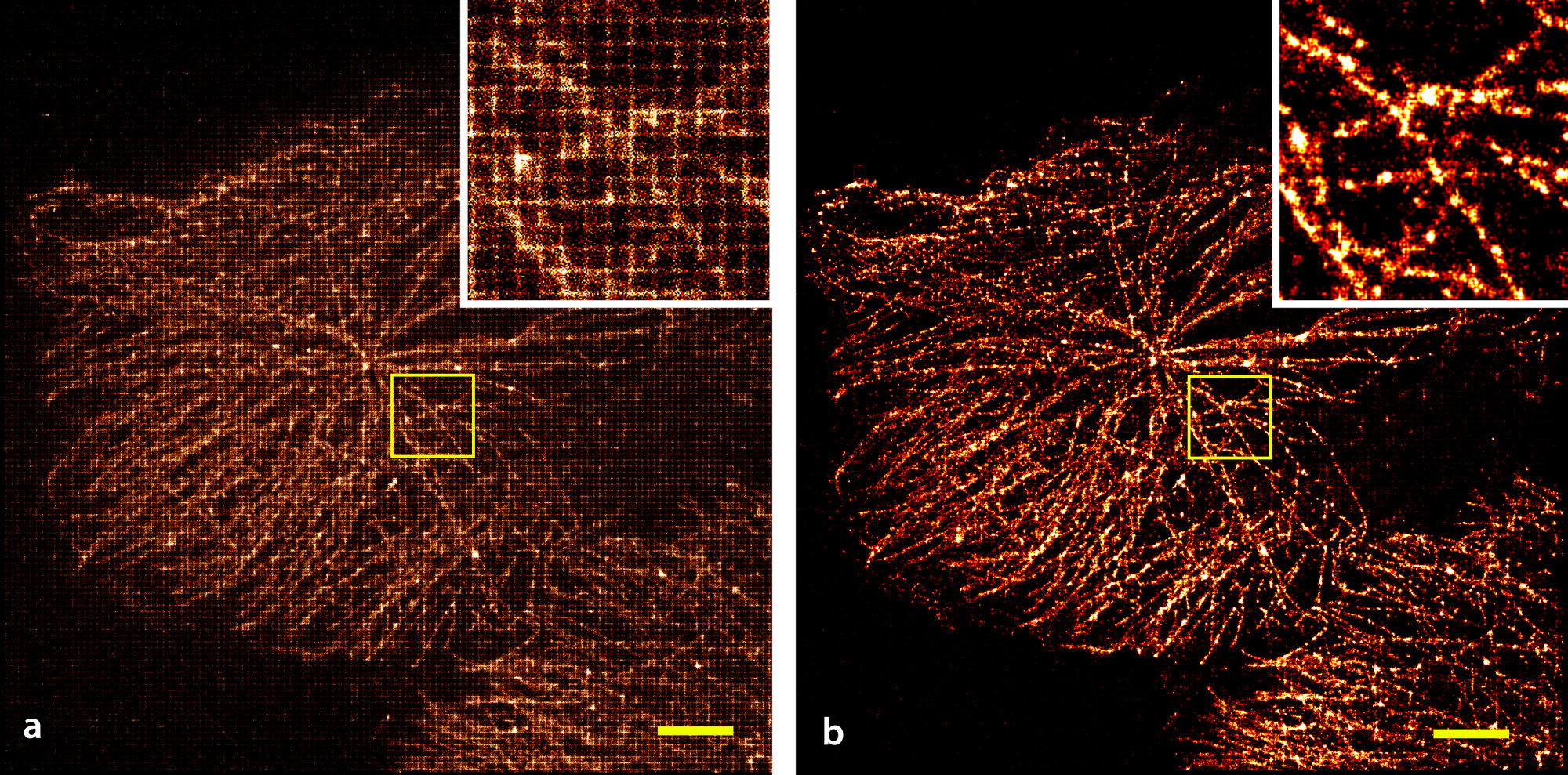}
	\caption[Fig 3:]{Comparison between unprocessed data and processed data. The neural network successfully got rid of the pattern effect which is due to the compression of the video stream. Scalebar = $1\,\mu m$}
	\label{fig:fig4}
\end{figure}
Particularly at low-light conditions, SNR decreased dramatically. The low signal does not only suffer a reduced dynamic range, but even is clipped to zero below a minimum intensity by the sensor and preprocessing hardware. Furthermore the video-compression algorithm encodes a specific block structure into the data, where the hard edges are interpreted as blinking fluorophores by the localization software Fig. (\ref{fig:fig4} a). Applying our learned forward model reduces such artifacts significantly (Fig. \ref{fig:fig4}b.). 

The NN was trained on a dataset consisting of frames from previously localized emitters from a noisy video-stream (see Supplement \ref{sec:datasetlocalized}) and artificially created blinking fluorophores processed by our simulated camera model (see Supplement \ref{sec:datasetsimu}).

In cases, where the imaging conditions are good, the NN is not necessarily optimizing the result. This method is thus especially useful for the wide variety of cameras whose SNR is much worse or the sample preparation leads to low intensity in the \textit{d}STORM experiment. 

A next neighbor analysis of the ground truth data of a simulated STORM data-stack with non-processed and preprocessed simulated data stacks with compression artefacts are showing the strength of our NN approach. Especially out-of focus or by the compression artifact artificially created localization events were successfully eliminated. Hence the average localization accuracy (calculated as the mean deviation from the ground truth) was dramatically improved (see section \ref{sec:nnimprovement}).

\section{Discussion}
We have demonstrated the suitability of using a modern smartphone camera for imaging beyond the diffraction limit. Even considering the limitations imposed by current smartphone hard- and software, i.e.  low-light performance and artifacts caused by compression and image “enhancement” algorithms currently unavoidably being introduced by the camera chips, we have been able to resolve sub-diffraction detail in cytoskeletal structures on a level similar to conventional \textit{d}STORM setups. 
\newline 
Preprocessing by a trained neural network albeit being useful in some cases has to be used with care, because one can hardly judge, what the network really learned. The localization uncertainty learned by misaligned data-pairs produces blurred results whose resolution is less then unprocessed data. Yet, it allowed a proof-of-concept recovery of the (d)STORM forward model for the smartphone sensor at poor imaging conditions. At the same time it may introduce additional localization uncertainties. 

\section{Conclusions }
We showed that widely available cellphone cameras can be used for SMLM yielding image quality comparable to much more expensive professional cameras. This is a milestone in the development of an overall very cost-effective SMLM system. It also paves the way of making super-resolution microscopy widely available, which has the potential to accelerate scientific work. Specific neuronal processing units inside new cellphones makes it also very interesting to integrate the localization done by a NN inside the acquisition device, reducing the footprint of this technique significantly. 
Education also benefits from this approach directly, where ordinary cellphones are readily available. This removes barriers for future research of all levels of society and could bring new contributions to the field of biological and medical research.   

\section{Acknowledgments}
We thank especially Ingo Fuchs who helped a lot in understanding the principles of the acquisition process of cellphone cameras. We also thank im for the support of the software design.
Our project was funded by the DFG Transregio Project TRR166, TP04.
\newpage

\appendix
\section*{Supplementary Material}
\addcontentsline{toc}{section}{Appendices}
\renewcommand{\thesubsection}{\Alph{subsection}}

\section{Hardware Setup}
\label{sec:hardset}
\begin{figure}[h!]
	\centering
	\includegraphics[width=0.6\linewidth]{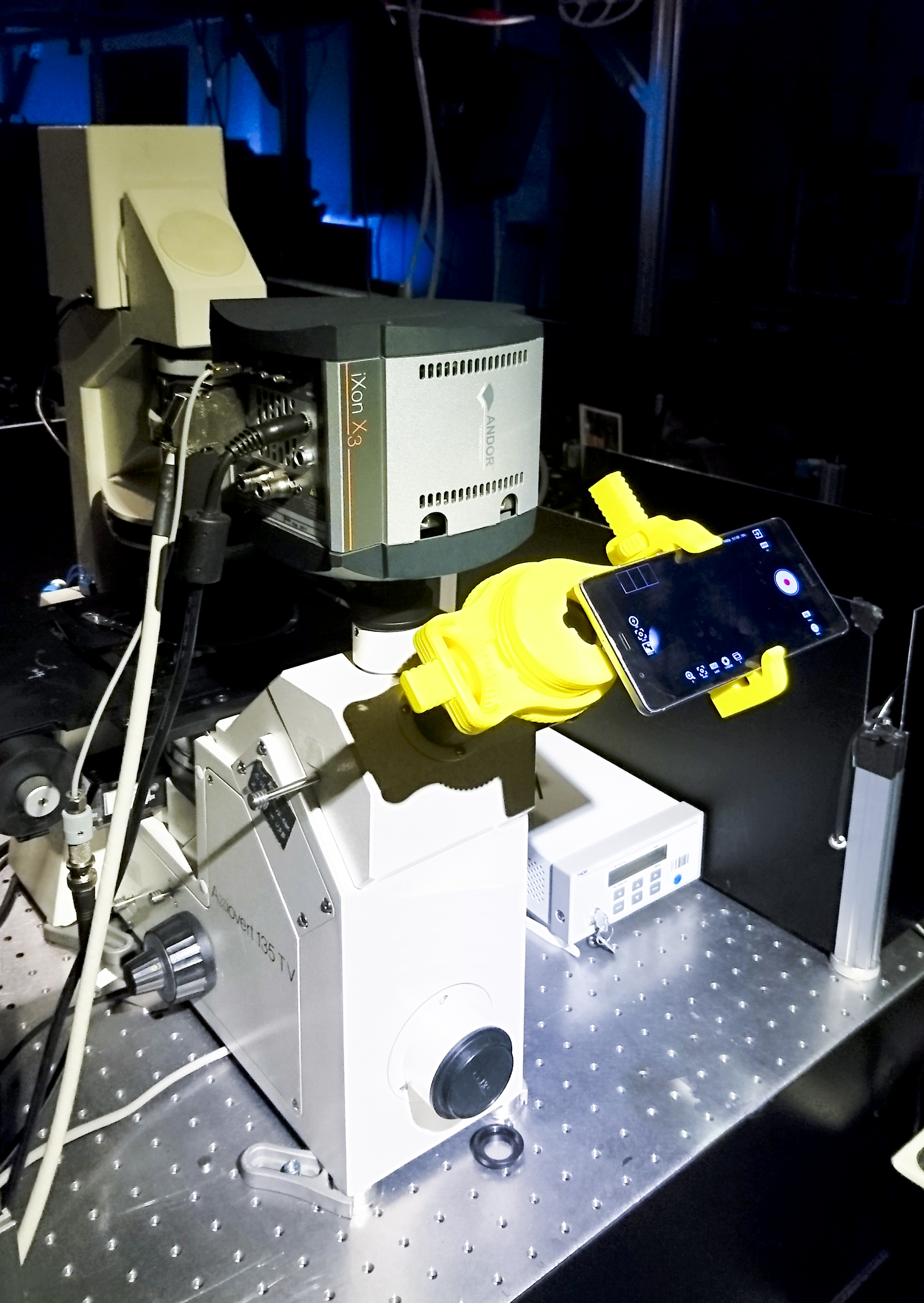}
	\caption[Fig 1:]{The custom made, 3D-printed cellphone adaptor in place on the Zeiss Axiovert microscope body }
	\label{fig:fig1}
\end{figure}

\section{Camera Characterisitics }
\label{sec:camchar}
\begin{figure}[h!]
	\centering
	\includegraphics[width=0.9\linewidth]{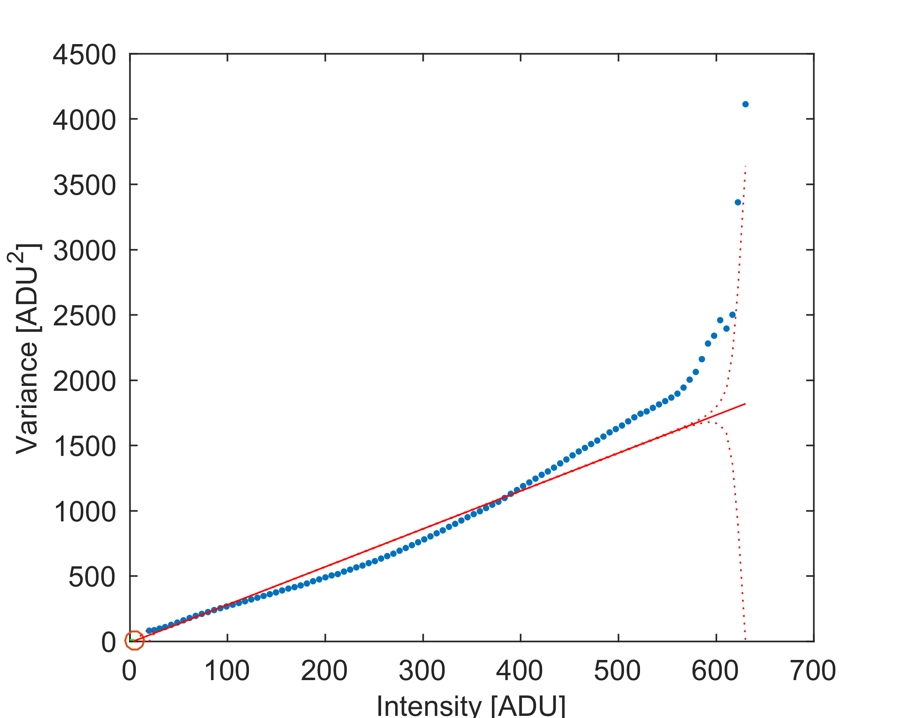}
	\caption[Fig 4:]{Mean-variance plot generated using a series (ISO=3200) of unprocessed raw images ("snap-mode") acquired by a HUAWEI P9 camera (blue points). The variance does not increase linear with the mean intensity (best fit in red), so that there must be additional noise sources in the images. Measured values for offset = 4,074 ADU; gain = 0,34 e/ADU; readnoise(Bg) = 1,23 e RMS; fixed pattern noise (BG) = 0.343 e RMS.}
	\label{fig:fig6}
\end{figure}

A characteristic mean-variance plot is generated for the HUAWEI P9 camera, from a set of unprocessed raw images aquired in snap-mode (12Bit DNG) of an intentionally defocussed but stationary object (see Fig. \ref{fig:fig6}) by using the Dip-Image \cite{Prof.Dr.BerndRieger2018} "cal$\_$readnoise" routine. It can be seen that the variance does not increase linearly with the mean intensity as it should for a shot-noise-limited sensor \cite{Stijns2011}. 
The noise parameters extracted from the low-intensity range of the curve are offset = 4,074 ADU; gain = 0,34 e/ADU; readnoise(Bg) = 1,23 $e^-$ RMS; at an ISO3200, which was also used during our measurements.

Especially noteworthy is the low readnoise in Fig. \ref{fig:fig6}. However, it cannot be guaranteed that the hardware-based preprocessing especially in the video-mode does not alter this value. 

\begin{figure}[h!]
	\centering
	\includegraphics[width=0.9\linewidth]{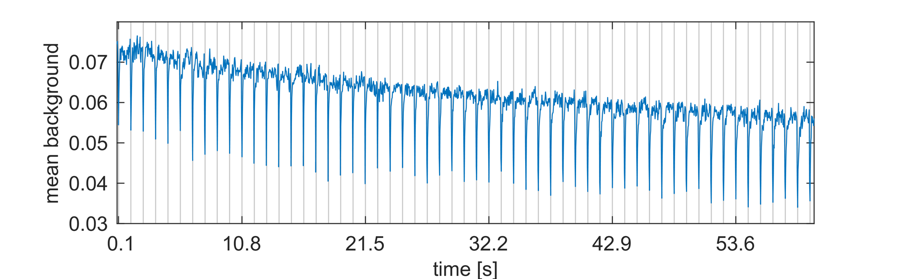}
	\caption[Fig 5:]{Mean of a dark image acquired with a HUAWEI P9 camera and compressed with a H264 encoder as a function over time. The overall decay of the signal and the periodic drop of the signal demonstrates the disadvantages of a compressed video signal. Equidistant gray lines (vertical) with 1.07\,s spacing show the periodicity of the drop in the intensity signal.}
	\label{fig:fig5}
\end{figure}

A series of images of dark background acquired in the video-mode with the HUAWEI P9, automatically compressed with a H264 encoder, shows another problematic property. The mean of each dark frame over time is shown in Fig. \ref{fig:fig5}. It can be seen that the overall intensity is reduced over time. This might be a thermal problem, although the signal is expected to rise rather than drop. This effect, however, might also be caused by the compression of the incoming signal. Unfortunatley the HUAWEI P9 has neither a temperature sensor on the chip nor a reproducible data compression, so that it stays unclear what the reason is. In addition, the signal drops periodically (every 1.07\,s at 20\,fps) which seems to be a compression artifact. A homogeneous although slightly noisy line would have been expected. The acquisition of a video shows already the drawback and limitations of the compression. 

Looking at Fig \ref{fig:fig5} it can be seen that the HUAWEI P9 sensor and video compression codec add a remarkable amount of noise to an image.
Imaging techniques that require extensive image processing will have problems using such data. However, wide-field microscopy of bright specimens will be less affected. The localization accuracy of SMLM will be impaired the more noise each raw image contains.
However noise allone will not create artifacts. The occurring drop of the overall intensity is no problem either, because each image is processed individually and offset variations are automatically accounted for by rapidSTORM and ThunderSTORM. 

\section{Neural Network Architecture and Training}
\label{sec:nnarchitecture}
\subsection{Generation of the training dataset}
\label{sec:nndataset}

Here we describe the generation of the dataset "\textit{A} to \textit{B}" to feed the modified version of the pix2pix generative adversary neuronal network detailed described in Supplement \ref{sec:networkarch}. The idea is to recover $B$, i.e. adjusted for the effects of compression or noise, from a captured video-frame \textit{A} by applying the trained network. We propose three different methods to produce suitable training sets.

\subsubsection{Dataset from camera’s model simulation}
\label{sec:datasetsimu}
We first created a data-stack of simulated STORM frames using the software testSTORM \cite{Novak2017}. Parameters of line- as well as randomly oriented vesicles structures were chosen according to our experimental conditions. In a later step we estimated a camera model, based on the properties determined in Supplement \ref{sec:camchar}, to introduce noise into the data, before they were compressed by the H.264 video-codec in MATLAB \cite{TheMathWorksInc.}. The compression ratio was tuned, so that the compression artifacts looked similar to the one from the original acquisition ("Video-Quality": $20\,\%$). Combining the decompressed simulated video-frames with the position of the ground truth events convolved with the system’s PSF represent the dataset for training the NN.
\newline
Following this procedure gives only an estimated forward model of the unknown camera black-box and therefore most probably does not account for all properties of the data acquired by the cell-phone camera.

\subsubsection{Dataset from imaged groundtruth data from a second display}
\label{sec:datasetsecondary}
To incorporate unknown influence of the camera's firmware, we created a dataset consisting of captured frames from ground truth data (see Supplement \ref{sec:datasetsimu}). The time-series was played as a movie on a secondary LCD from another cellphone (Sony Xperia Z5, 1080p, IPS-LCD, Japan) and is recorded in the video-mode by the HUAWEI P9 at about 114 mm distance. Experimental parameters like exposure-time and sensitivity (ISO), as well as the expected PSF (point spread function) are chosen to correspond to the \textit{d}STORM experiments. To avoid a moiré-effect due to the pixelated structure of the ground-truth display, we slightly defocused the image on the HUAWEI P9. Inhomogeneities in the illumination were compensated by flat-fielding the images before they were fed to the NN for training. 
\newline
To produce a matching dataset, where each camera frame corresponds to its equivalent ground-truth (GT) frame displayed on the secondary screen, we encoded a barcode representing the actual frame number of the ground truth data, into the displayed video. In a post-processing step, first the barcode is readout and assigned to the equivalent ground-truth frame number. To avoid any ambiguous overlap exposures between any two frames, we took only every 3rd frame after detecting a barcode. The framerate of the GT-movie and acquisition device was set to have a correct sampling in time (e.g. four frames).

To account for misalignment between the two frames in the registration procedure, we first estimated the affine transformation parameters and later transformed the GT frames using the Dip-Image toolbox \cite{Prof.Dr.BerndRieger2018} for MATLAB \cite{TheMathWorksInc.} to have sub-pixel accurate registered images.

\subsubsection{Dataset from localized dSTORM data}
\label{sec:datasetlocalized}
Due to barrel-like lens distortions, the frames in the previous method (section \ref{sec:datasetsecondary}) did not match over the entire field of view. Thus the NN learned a location invariant localization uncertainty. The U-NET \cite{Ronneberger2015}  based generator of the GAN recovers the PSF at randomly shifted positions, based on the misaligned dataset in the range of about a quarter of a pixel. When localizing the processed image-series using ThunderSTORM the randomly varying shift of the PSF produces blurry results.

Therefore we created a third way to generate a dataset by taking captured \textit{d}STORM time-series  using the video-mode from real biological cells (labelled microtubules) under optimal conditions. After localizing the blinking events using ThunderSTORM for each frame, we extract the detected emitters and draw gaussian PSFs according to the fluorophore's position on a bitmap image. 
\newline
To not only learn the forward model from the ThunderSTORM PSF-fitting algorithm, we also incorporated data from the method described in Supplement \ref{sec:datasetsimu}. This enhanced the quality after processing the frames and increases the number of correctly detected blinking events. 
It also successfully accounted for variations in sample’s background as well as in the camera parameters. 
\subsection{Network Architecture }
\label{sec:networkarch}
The registered (time/location) data-pairs feed our modified version of the image-to-image GAN network, which was implemented in the open-source ML library PyTorch \cite{Paszke2017}. The code is based on the open-sourced version described in \cite{Zhu2018}.
\newline
To circumvent a checkerboard-like artifact in the reconstruction process, we replaced the deconvolution operation in the decoding step of the U-NET \cite{Ronneberger2015} generator by a resize-convolution layer as suggested in \cite{Odena2016}, based on the nearest-neighborhood interpolation method. This immediately eliminates the high-frequency patterns due to the low coverage of the convolutions in the deconvolution process.

The neural network (NN) was trained on a Nvidia Titan X GPU over 5000 frames with equal acquisition parameters like the one in the real \textit{d}STORM experiment taken from method one (Supplement \ref{sec:datasetsimu}) and three (Supplement \ref{sec:datasetlocalized}) to equal parts. We use minibatch SGD and relied on the ADAM optimization scheme with learning rate of $1E-4$ and momentum of beta\,$= 0.25$. The number of first-layer filters in the PatchGAN-discriminator \cite{Hou2015} and U-NET generator was chosen to be $32$ each. 
\newline
Compared to approaches like the recent published Deep-STORM by Nehme et al. \cite{Nehme2018}, the here presented GAN does not require a particular cost-function which tries to produce optimal result with the generator. Instead the discriminator tries to distinguish whether the results are coming from the generator or provided by the GT data. Hence the GAN should come up with a learned forward model which successfully includes all unknown effects to map the acquired camera frames to images which then follow the forward model of the \textit{d}STORM technique much better. This facilitates a parameter free optimization technique. 

\subsubsection{Training and Evaluating the Network }
Our experiment showed, that the training converged to equilibrium after 100 epochs at a batchsize of 4 frames, which follows in about 2h time-effort on an ordinary desktop machine with 64Gb RAM, Intel Xeon octacore, and a Nvidia TitanX graphics card with 12GB memory. It is worth noting, that a precise alignment of the data is crucial, otherwise the recovered events will be shifted by an unknown amount and the localization fails due to smeared-out blinking events. 
\newline
Applying the trained network takes about 5 Minutes for a stack of fifteen thousand frames of 256$\times$256 pixels. Due to the convolutional-architecture of the PatchGAN it is possible to process data with framesizes different than the training dataset in case the pixel information in the radius of the convolutional kernel are uncorrelated i.e. following random Markovian fields \cite{Li2016, Isola2016b}, which is the case for \textit{d}STORM measurements. 
Our approach does not rely on any specific class of imaged objects, nor does it need any parameters other than a dataset which mimics the experimental data in the sense of acquisition parameters. 

\subsubsection{Quantitative Analysis of results using NN}
\label{sec:nnimprovement}

\begin{figure}[h!]
	\centering
	\includegraphics[width=1\linewidth]{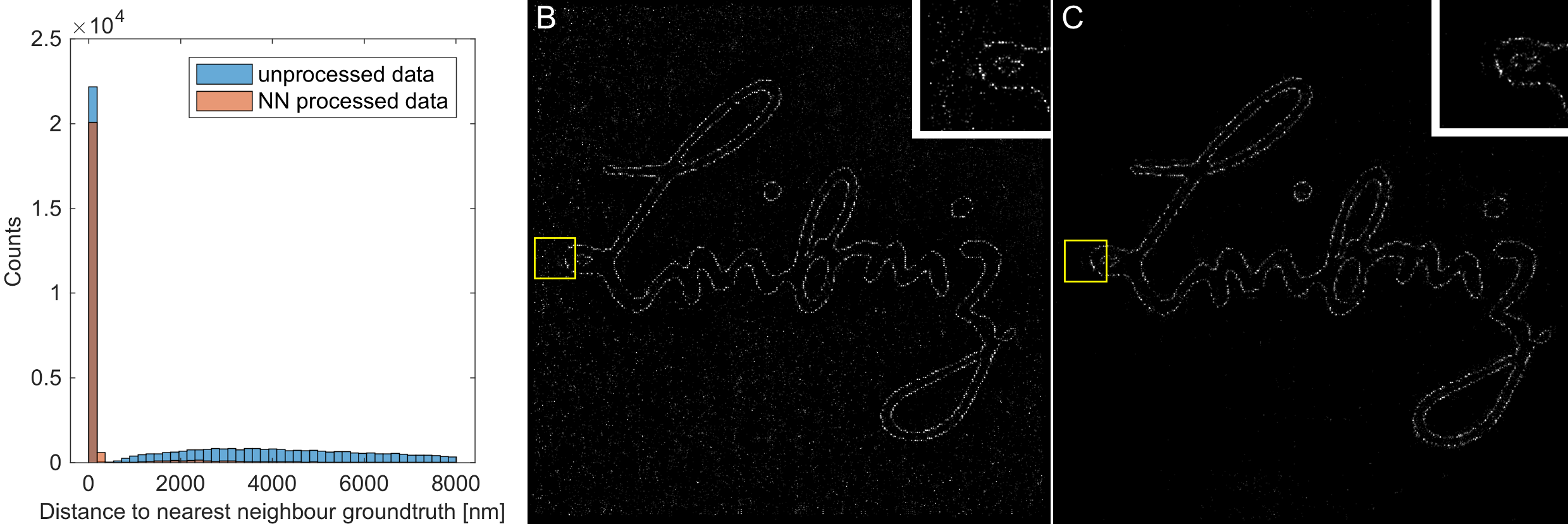}
	\caption[Fig 7:]{Histogram in a, where we compare the distance between the ground-truth positions and the detected localization from unprocessed and NN-processed data. From this and the reconstructed SMLM images in b (unprocessed data) and c (NN-processed data) it is clearly visible, that the NN successfully distinguish between true blinking events and detected events coming from noise- or compression artifacts. Number of detected events compared to ground-truth: 22658; unprocssed video: 50509; NN-processed video: 22658}
	\label{fig:fig7}
\end{figure}

To give a quantitative measure of how well the NN recovers data suffering from noise and compression artifacts we again simulate a STORM-dataset from the Lebniz-Institute's logo (Fig. \ref{fig:fig7} b, c) using testSTORM based on our camera model proposed in \ref{sec:camchar}. We localize the events using ThunderSTORM and measure the distance between recovered blinking and ground-truth positions using MATLAB's function "knnsearch". 

Especially for experiments, where noise is dominant, the NN is capable to differentiate between false-positive and true blinking events. From the histogram in Fig. \ref{fig:fig7}, giving the distances between nearest neighbors in GT and unprocessed as well as GT and NN-processed events for each frame, it is clearly visible, that the NN enhances the localization accuracy.

\printbibliography



\end{document}